\documentclass[aps, twocolumn, superscriptaddress, amsmath, amssymb, floatfix, prb, longbibliography]{revtex4-2}

\usepackage{amsfonts}
\usepackage{graphicx}
\usepackage{enumerate}
\usepackage{bm}
\usepackage{bbm}
\usepackage{color}
\usepackage[dvipsnames]{xcolor}
\definecolor{prlblue}{rgb}{0.18,0.19,0.57}
\usepackage{braket}
\usepackage{tikz}
\usepackage{ulem}
\usetikzlibrary{patterns.meta} 
\usetikzlibrary{calc}
\usetikzlibrary{shapes.geometric}

\makeatletter
\newsavebox{\@brx}
\newcommand{\llangle}[1][]{\savebox{\@brx}{\(\m@th{#1\langle}\)}%
  \mathopen{\copy\@brx\kern-0.5\wd\@brx\usebox{\@brx}}}
\newcommand{\rrangle}[1][]{\savebox{\@brx}{\(\m@th{#1\rangle}\)}%
  \mathclose{\copy\@brx\kern-0.5\wd\@brx\usebox{\@brx}}}
\makeatother

\newcommand{\E}{\mathcal{E}}
\newcommand{\Z}{\mathbb{Z}}

\newcommand{\ii}{\mathrm{i}}
\newcommand{\tr}{\mathrm{Tr}}
\newcommand{\be}{\begin{equation}}
\newcommand{\ee}{\end{equation}}
\usepackage{hyperref}
\usepackage{natbib}
\begin{document}

\title{Measures of Chirality in Mixed-State Topological Phases}

\author{Shijun Sun}
\affiliation{School of Physics, Georgia Institute of Technology, Atlanta, Georgia 30332, USA}

\author{Rasmit Devkota}
\affiliation{School of Physics, Georgia Institute of Technology, Atlanta, Georgia 30332, USA}
\affiliation{School of Mathematics, Georgia Institute of Technology, Atlanta, Georgia 30332, USA}

\author{Bader Aldossari}
\affiliation{School of Physics, Georgia Institute of Technology, Atlanta, Georgia 30332, USA}
\affiliation{Physics Department, King Fahd University of Petroleum and Minerals, Dhahran, Saudi Arabia}

\author{Zhu-Xi Luo}
\affiliation{School of Physics, Georgia Institute of Technology, Atlanta, Georgia 30332, USA}

\date{\today}

\begin{abstract}
What does it mean for a mixed-state topological phase to be chiral? Mathematically, chirality can be sharply characterized through the symmetry algebra of the mixed state. Physically, however, the question is far more subtle. In pure states, chirality in topological phase is tied to a web of familiar diagnostics, involving bulk-boundary correspondence, a gapless entanglement spectrum, a nontrivial modular commutator, and a quantized thermal Hall response. We show that none of these diagnostics remain  reliable in mixed states. Instead, for decohered topological phases with a known error-free parent state, we propose two relative-entropy-based measures that can diagnose chirality, with one of them further extracting the chiral central charge. As a concrete example, we use these measures to analyze the $em^2$-decohered $\mathbb{Z}_3$ toric code, identifying the chirality transition to be in the random-bond Potts class for the particle-hole symmetric channel, and the polarized random bond Potts class for asymmetric channels. Our results emphasize how mixed-state topology demands intrinsically new diagnostics beyond direct analogues of pure-state probes.
\end{abstract}

\maketitle

\setcounter{tocdepth}{1} 
{
  \hypersetup{linkcolor=magenta}
  \tableofcontents
}

\section{Introduction}
\label{sec:intro}

Topological phases of matter have profoundly reshaped our understanding of quantum phases of matter in pure states \cite{doi:10.1142/S0217979290000139,Wen_2017,Kitaev_2006}. They are now reappearing at the frontier of the subject in a new form: as phases of mixed states arising from noise, such as decoherence, dissipation, measurements, and disorder.  Recent work has revealed broad classes of intrinsically mixed-state topological phases, whose universal properties need not have a pure-state ground-state counterpart \cite{PRXQuantum.6.010314,zini2021mixedstatetqfts,PRXQuantum.6.010313,PRXQuantum.6.010315}.  Among them, chiral mixed-state phases are especially intriguing: Like their pure-state counterparts, they possess chiral anyon statistics,
chiral symmetry algebras, and consequently nontrivial chiral central charge.  Yet, in sharp contrast with pure chiral topological orders, they can admit commuting-projector density-matrix representations on the lattice \cite{Haah_2023}, a
possibility forbidden for pure chiral phases \cite{Kapustin_2020}.  This tension raises a natural question: what physical signatures of chirality are robust once the state is mixed?

The pressing need to answer this question is both fundamental and practical. As of now, the vast landscape of mixed-state topological orders remains largely unexplored beyond abstract categorical formulations \cite{Dennis_2002,PRXQuantum.6.010314,zini2021mixedstatetqfts,PRXQuantum.6.010313,PRXQuantum.6.010315,ogata2025mixedstatetopologicalorder,KIKUCHI_2025}. Existing studies mostly focused on \(\mathbb Z_2\)
topological orders under a variety of physically motivated noises, including for example incoherent errors \cite{Dennis_2002,PRXQuantum.5.020343,Bao_2026,PRXQuantum.6.010314,Chen_2024}, coherent errors  \cite{PhysRevB.110.125152,Lee_2025,wang2025decoherenceinducedselfdualcriticalitytopological}, disorders \cite{PhysRevB.83.075124, PhysRevA.85.022313, PhysRevB.111.115137} and random measurements \cite{Lavasani_2021,Negari_2024,PhysRevB.109.224306}. Simple (pure $Z$ or pure $X$) decoherence channels in CSS codes \cite{Su_2024,lyons2024understandingstabilizercodeslocal} have also been examined. However, these setups do not allow for chirality transitions, and therefore cannot capture the qualitatively new physics associated with chiral anyon braiding and potential gravitational anomalies \cite{ALVAREZGAUME1984269}. To understand the latter, it is therefore essential to examine more general mixed-state topological phases. From the perspective of fault-tolerant quantum computation, understanding mixed-state chirality is also a key step: for example, the simplest topological order supporting universal topological quantum computation is the chiral Fibonacci order \cite{freedman2002modular,RevModPhys.80.1083}. Moreover, in physical platforms where dissipationless chiral edge transport is a key resource \cite{PhysRevB.25.2185,PhysRevB.55.15832}, understanding how chirality behaves in an inevitably open environment is of direct importance.

In this work we approach this issue by revisiting several standard and interconnected signatures of pure-state chirality, including gapless boundary theory \cite{ALVAREZGAUME1984269,hellerman2021,Fan_2022},
gapless entanglement spectrum \cite{Li_2008, PhysRevLett.108.196402}, and nontrivial modular commutator \cite{Kim_2022_0,Fan_2022,Kim_2022,vardhan2025}. We show that, in mixed-state chiral
topological phases, none of are generically reliable. 
This breakdown is itself instructive: it shows that mixed-state chirality is not merely a noisy remnant of pure-state chirality, but an intrinsically open-system notion requiring diagnostics adapted to decoherence and
information loss. We further propose two quantum-information-theoretic chirality measures based on relative entropy \cite{10.2996/kmj/1138844604}.  For decohered topological
phases with a known error-free parent state, these measures diagnose whether chirality survives the noise, and one of them can quantitatively extract the chiral central charge. 
As a proof of concept, we analyze in detail the $em^2$-decohered $\mathbb{Z}_3$ topological order, and map the relative-entropy diagnostic to correlators in random-bond three-state Potts models and its variants. While  particle-hole symmetric decoherence leads to the conventional random bond Potts model on the Nishimori line, particle-hole asymmetric decoherence, where the $em^2$ and $e^2m$ anyons are proliferated with different probabilities, leading to a polarized random-bond Potts model on a generalized Nishimori surface. 
Remarkably, despite the chirality of the decoherence channel, the associated critical theory is expected to be described by a nonchiral CFT. 
These results provide the first steps toward a diagnostic framework for intrinsically mixed-state chirality.

\subsection{Preparations}

We start by recalling the algebraic setting that underlies these diagnostics. The intrinsically mixed-state topological phases of interest exhibit strong-to-trivial breaking of 1-form symmetries \cite{PhysRevB.111.115137}, each corresponding to an anyon type in the phase. The algebraic theory describing anyons in mixed states is that of the premodular tensor category \cite{PRXQuantum.6.010314,zini2021mixedstatetqfts,PRXQuantum.6.010313,PRXQuantum.6.010315}, which allows for non-vacuum anyons to have trivial statistics. 
For an anyon $a$ in the premodular tensor category, its quantum dimension $d_a$ is a real number, and intrinsic topological spin $\theta_a$ a phase factor. By the Gauss-Milgram formula \cite{Kitaev_2006}, they are related to the chiral central charge $c_-$ via
\be
e^{2\pi \ii c_-/8} =\frac{1}{\mathcal{D}} \sum_a d_a^2 \theta_a. 
\label{eq:c}
\ee
Here $\mathcal{D}=(\sum_a d_a^2)^{1/2}$ is the total quantum dimension. Since stacking with an invertible bosonic $E_8$ state shifts $c_-$ by $8$ without changing the anyon content, the above formula determines
$c_-$ only modulo $8$. Nontrivial chiral central charge is a sufficient but not necessary requirement for chirality. Recently, higher central charges $\xi_{\alpha}$ have been proposed \cite{Ng_2019,Ng_2022, Kaidi_2022} to capture chirality-related anomalies invisible to standard chiral central charge. For certain Abelian topological phases, $\xi_{\alpha}\propto \sum_a d_a^2 \theta_a^{\alpha}$.

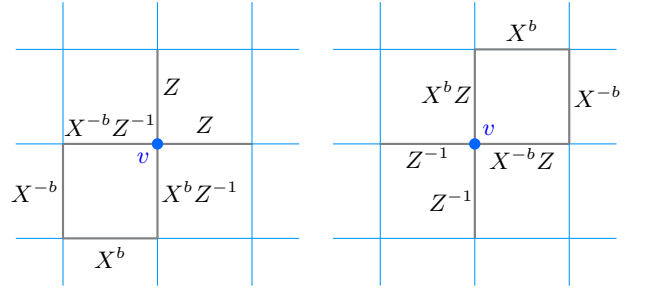
\begin{figure}[htbp]
\centering
\begin{tikzpicture}[scale=1.25]
\tikzset{grid line/.style={thin, color=blue!40!cyan}}
\draw[grid line] (-2.5,-1.5) grid (0.5,1.5);
\draw[thick,gray] (0-1,-1)--(0-1,1);
\draw[thick,gray] (-1-1,0)--(1-1,0);
\draw[thick,gray] (0-1,-1)--(-1-1,-1)--(-1-1,0);
\coordinate (v) at (0-1,0);
\fill[color=blue!60!cyan] (v) circle (0.06);
\node at (-1+0.15,0.6) {$Z$};
\node at (-0.5-1, 0.2) {$X^{-b}Z^{-1} $};
\node at (0.5-1,0.2) {$Z$};
\node at (-1+0.45,-0.5) {$X^bZ^{-1}$};
\node at (-0.5-1,-1.2) {$X^b$};
\node at (-2-0.3, -0.5) {$X^{-b}$};
\node at (-0.15-1,-0.15) {\textcolor{blue}{$v$}};
\end{tikzpicture}
\quad
\begin{tikzpicture}[scale=1.25]
\tikzset{grid line/.style={thin, color=blue!40!cyan}}
\draw[grid line] (-2.5,-1.5) grid (0.5,1.5);
\draw[thick,gray] (-1,-1)--(-1,1);
\draw[thick,gray] (-2,0)--(0,0);
\draw[thick,gray] (0,0)--(0,1)--(-1,1);
\coordinate (v) at (0-1,0);
\fill[color=blue!60!cyan] (v) circle (0.06);
\node at (-0.3-1,0.55) {$X^bZ$};
\node at (-0.5-1, -0.15) {$Z^{-1}$};
\node at (0.5-1,-0.15) {$X^{-b}Z$};
\node at (-0.25-1,-0.6) {$Z^{-1}$};
\node at (-0.5,1.2) {$X^b$};
\node at (0.3, 0.5) {$X^{-b}$};
\node at (0.15-1,0.15) {\textcolor{blue}{$v$}};
\end{tikzpicture}
\caption{Definition of Hamiltonian terms in $\mathbb{Z}_N$ toric code. Left: $O_v$. Right: $O_v'$. The definition depends on an integer $b$ which is a $\Z_N$ integer.}
\label{fig:def_O}
\end{figure}

Throughout this paper, we will use the $\Z_N$ toric code as an example parent topological phase, and the mixed-state topological phases can be obtained from the parent phase via anyon condensation \cite{Ellison_2022} and decoherence. General Abelian mixed-state topological orders follow by replacing \(\mathbb Z_N\) with products of cyclic groups. The degrees of freedom are $\Z_N$ qudits living on the links of a square lattice. The elementary operators are generalized Pauli operators satisfying $ZX=\omega XZ$, where $\omega=e^{2\pi \ii/N}$ is the $N^\text{th}$ root of unity. Focusing on odd $N$, we define the Hamiltonian terms  in an unconventional way to highlight its potential algebraic factorization into chiral and anti-chiral theories. Instead of using the usual star and plaquette operators, we will combine them into two types of  operators $O_v$ and $O_v'$ shown in fig. \ref{fig:def_O}, which depend on an integer $b$ satisfying $\gcd (N,b)=1$.  
The Hamiltonian is then a summation of locally commuting projectors which project into the space of $O_v=O_v'=1$, 
\be
H=-\frac{1}{N}\sum_v \sum_{a=0}^{N-1} (O_v^a + O_v^{\prime a}).
\label{eq:Ham}
\ee
The anyon excitations are labeled by $e^x m^y$ with quantum dimension $d_{e^xm^y}=1$ and topological spin $\theta_{e^xm^y}=\omega^{xy}$, where $x, y\in [0,N-1]$ are integers. The $O_v$ operator represents a closed loop of {$em^{-b}$} anyon, while the $O_v'$ operator represents a closed loop of the {$em^b$} anyon. One can break such loops open, and the elementary open string operators which create pair of $(em^{-b},e^{-1}m^{b})$ and pair of $(em^{b},e^{-1}m^{-b})$ commute with each other. 
In appendix \ref{app:setup}, we show this formulation of the toric code Hamiltonian is equivalent to the conventional definition. We will further define the logical operators $W_x, W_y, W_x', W_y'$ winding around non-contractible $x, y$ loops of the torus.

Below we will first introduce the two proposed diagnostics, and then discuss the pure state intuitions that fail for mixed states.

\section{Chiral central charge from relative entropy}
\label{sec:rel_S}

We first introduce a diagnostic of chiral central charge in a mixed-state topological phase obtained from decohering a known parent pure-state topological order.  This diagnostic does not require prior knowledge of the decoherence channel. From  \eqref{eq:c}, if the decoherence proliferates a subset of anyons, then the remaining mixed-state topological order will have a different set of surviving anyons and therefore a potentially different chiral central charge compared with the parent state. The chiral transition is then expected to coincide with the decodability transition of precisely those proliferated anyons. The latter can be diagnosed using relative entropy. Let $L_a(\ell)$ be an open string operator creating a pair of anyon $a$ and its antiparticle $\bar{a}$ at the endpoints of the long path $\ell$. Starting from the parent state $\rho_0$, we first compare the decohered state $\rho_\mu = \E_\mu [\rho_0]$ at error strength $\mu$ with the state $\rho_{a,\mu} = \E_\mu [L_a(\ell)\rho_0 L^\dagger_a(\ell)]$ obtained by first inserting the open string and then applying the same error channel. The difference between these two states is measured by the quantum relative entropy \cite{Umegaki1962, PRXQuantum.5.020343},
$
D(\rho_\mu||\rho_{a,\mu}) = \tr \left(\rho_\mu \log \rho_\mu\right) - \tr \left(\rho_\mu \log \rho_{a,\mu}\right).
$
If the anyon $a$ remains a well-defined excitation of the mixed-state topological order, then the open string should decay exponentially, reflected in the relative entropy as $D(\rho_\mu||\rho_{a,\mu})\propto |\ell|$ and diverges in the limit of large $\ell$. In contrast, if $a$ is proliferated, then the open-string can evaluate to order one, such that the relative entropy stays order one even in the thermodynamic limit. 
So the scaling of the relative entropy can then be used as an indicator of which anyons survive the error channel. One can define a binary characterization of the survival of anyon $a$: 
\be
\chi_{a}(\mu) = \Theta\left( \lim_{|\ell|\rightarrow\infty} \frac{1}{|\ell|} D(\rho_\mu||\rho_{a,\mu}) \right),
\ee
where $\Theta$ is the Heaviside theta function satisfying $\Theta(x)=1$ for $x>0$ and $\Theta(0)=0$.  Thus $\chi_a(\mu)=1$ if $a$ survives, and $\chi_a(\mu)=0$ if $a$ proliferates. The chiral central charge of the remaining mixed-state topological order can then be extracted through the topological twist of the surviving anyons, 
\be
c_- (\mu) = \frac{4}{\pi}\arg \left(\sum_a d^2_a \theta_a \chi_{a}(\mu)\right),
\label{eq:c_D}
\ee
where the summation in the first line runs over all anyon types in the parent topological order.

We now examine a class of examples where the parent theory is the $\mathbb{Z}_N$ toric code with odd $N$. Odd $N$ is chosen because the parent topological order allows for factorization into two topological orders corresponding to the stabilizer families  $O_v$ and $O'_v$ defined in fig.\ \ref{fig:def_O}. Without decoherence, both families are present and the topological spins sum to a real number, leading to zero chiral central charge as expected. Suppose the channel decoheres out the {$O_v$} family such that the indicator $\chi_{em^{-b}}(\infty)$ vanishes. The {$O_v'$} family is left intact, with $\chi_{em^b}(\infty)=1$. Then \eqref{eq:c_D} yields
\begin{equation}
    c_- (\infty)
    = \frac{4}{\pi} \arg\left(\sum_{k=0}^{N-1} \omega^{ b k^2}\right).
\end{equation}
For $N=3$ and $b=\pm 1$, this gives $c_- (\infty) = \pm 2$. For $N=5$, whether there is chirality depends on the choice of $b$: when $b=\pm 1$, $c_-(\infty)=0$; when $b=\pm 2$, $c_-(\infty)=\pm 4$. These are consistent with the expectations from modular tensor category.

The chiral transition point can be obtained from the replica limit of the Rényi-$n$ relative entropy.  For the particle-hole symmetric channel at $N=3$ and $b=1$, this limit maps the relative entropy to the disorder-averaged excess free energy of a defect line in the random-bond three-state Potts model on the Nishimori line~\cite{Nishimori1983,Jacobsen_2002}.  The corresponding transition occurs at the critical decoherence strength $\mu_c\simeq0.2682$.

This analysis extends naturally to particle-hole asymmetric channels, for which the two nontrivial error processes occur with unequal probabilities. At $N=3$, the replica limit instead gives a polarized random-bond three-state Potts model on a generalized Nishimori surface.  Although its critical line is not presently known, the Nishimori--Nemoto conjecture~\cite{Nishimori_2002} gives $\mu_c^{\mathrm{conj.}}\simeq0.2923$ at $x=0.5$.  Details of both mappings are given in Appendix~\ref{app:Potts}.

We note that higher central charges $\xi_{\alpha}$ can also be obtained using the aforementioned method. When $\gcd (\alpha, N)=1$, one can obtain the higher central charge simply by replacing $\theta_a$ by $\theta_a^{\alpha}$ in \eqref{eq:c_D} and modifying the overall normalization factor.

\section{Order parameter for Time Reversal Breaking}
\label{sec:OP}

Chirality reverses under time reversal. It is therefore natural to detect chirality via order parameter of the time reversal symmetry. However, there are two subtleties: The first one is that the action of time reversal on the microscopic lattice operators is not always clear, and can vary even within the same phase. In other words, time reversal here is a topological operation that does not necessarily coincide with the time reversal of physical degrees of freedom, and can generically permute anyon types. This subtlety is already present in pure state case. Another subtlety intrinsic to mixed state physics is that the order parameter cannot be naïvely evaluated as a linear observable; to detect potential phase transitions, correlators nonlinear in the density matrix are often involved. Since both subtleties are irrelevant at exactly solvable fixed points, below we will begin by examining these special points, and then extend beyond them. 

Suppose the (topological) time reversal symmetry $\mathcal{T}$ sends anyon $a$ to $a^*$. Denote $\gamma$ as a closed, oriented loop on the lattice, and $W_a(\gamma)$  the oriented loop operator for anyon $a$ (or, in the Abelian case, the 1-form symmetry operator) defined on $\gamma$. Then the order parameter for time reversal breaking can be chosen as 
\be
\lim_{|\gamma|\rightarrow \infty} \langle W_a (\gamma) - \mathcal{T} W_a (\gamma) \mathcal{T}^{-1}\rangle=\lim_{|\gamma|\rightarrow \infty} \langle W_a (\gamma) - W_{a^*} (\gamma) \rangle,
\label{eq:OP}
\ee
where for now we view the expectation value $\langle\ \cdot\ \rangle$ as evaluated in $\tr [\rho \ \cdot\ ]$. If the topological phase is nonchiral, by definition the order parameter must vanish. 

Take the simple example of the decoherence of the $em^{-1}$ anyon, or the anti-chiral topological sector, from the $\Z_3$ toric code. At zero decoherence strength, the system lives in the nonchiral topological phase, while at maximal decoherence strength it's in the chiral $\Z_3^{(1)}$ topological phase. Choosing {$a=em^{-1}$} such that {$a^*=em$}, then $W_a (\partial R) =\prod_{v\in R} O_v$ and $W_{a^*}(\partial R)=\prod_{v\in R} O_v'$, where $R$ is a region such that its boundary is the closed $\gamma$ loop. It is easy to evaluate that $\langle W_{a^*}\rangle=1$ regardless of decoherence strength, because the decoherence channel leaves the chiral sector intact. In contrast, $W_{a}$ evaluates to $1$ without decoherence, and $0$ at maximal decoherence. Therefore, the order parameter can correctly predict chirality at these extreme points. 

Straightforward extension of this definition beyond fixed points will fail -- for instance, even in the case where the anti-chiral sector $\{a\}$ is fully decohered out, those destroyed topological symmetries $\langle W_{a}(\gamma)\rangle$ are still expected to scale as perimeter law $e^{-c_1|\gamma|}$ (instead of area law $e^{-c_2|\gamma|^2}$), except at the fixed point \cite{PhysRevD.19.3682,PRXQuantum.5.020343}. Here $c_i$'s are constants which keep the exponents dimensionless. The leftover intact chiral sector $\langle W_{a^*}(\gamma)\rangle$ will also scale as perimeter law away from fixed point, therefore the order parameter defined in \eqref{eq:OP} is not capable of pinning down the chirality conclusively. 

To attack this problem, one must twist the definition in \eqref{eq:OP}: we will work with nonlinear expectation values and replace the closed loops in the equation by open strings. Our definition of the modified order parameter $J (\E, \rho_0, L_a (\ell), L_{a*}(\ell) )$ is:
\begin{align}
J=\ & \tr \left( \E [\rho_0] 
\big( \log \E [L_a (\ell)\rho_0 L_a^{\dagger} (\ell)]\right.\nonumber\\
& \qquad \qquad\left. - \log \E [L_{a^*} (\ell)\rho_0 L_{a^*}^{\dagger} (\ell)]\big) \right).
\label{eq:J}
\end{align}
Here $\ell$ is again an open string, and $L_a(\ell)$ is the operator which creates a pair of $a$ and $\bar{a}$ excitations at the endpoints of $\ell$. (Notice that in general the antiparticle $\bar{a}$ of $a$ is not necessarily equal to the time reversal partner $a^*$ of $a$.) 
Equation \eqref{eq:J} can be viewed as the difference between two relative entropies \footnote{The more naïve choice of the relative entropy between $\E [L_a (\ell)\rho_0 L_a^{\dagger} (\ell)]$ and $ \E [L_{a^*} (\ell)\rho_0 L_{a^*}^{\dagger} (\ell)]$ will not be able to detect the phase transition, because in both phases, the two states are always distinguishable.}, and computed following the same steps as in the previous section \ref{sec:rel_S}.

Comparing withthe measure in the previous section, the calculation of $J$ does not require the knowledge of the full set of open string operators for each anyon type before decoherence. For fixed $\E$ and $\rho_0$, as long as there exists one pair of $(a,a^*)$ such that $J$ is nonzero, then the mixed-state topological phase is chiral. However, it is not able to extract the value of chiral central charge.

\section{Garden of bad but instructive measures}
\label{sec:failure}

This section collects several conceptually natural diagnostics inspired by their pure-state counterparts, which nevertheless fail at different stages. We view these failures as highly instructive: they clarify the obstacles to defining chirality in mixed states and may point toward further modifications that ultimately lead to viable diagnostics.

\subsection{Modular shear}
\label{subsec:shear}

We begin with a diagnostic that is (1) highly effective at fixed points, but (2) fails to reproduce the correct phase boundaries and (3) does not admit an easy remedy. Compared with the diagnostics proposed in the previous
sections, it can nevertheless be useful in a more limited setting: once a fixed point has been reached, for example through renormalization or tensor
network methods, it may provide a quick check of chirality.

The topological spin $\theta_a$ in \eqref{eq:c} is associated with the modular
$T$ transformation on the torus, the large diffeomorphism that twists
one noncontractible cycle of the torus once around the other. For a two-dimensional torus with continuous Cartesian coordinates $(x,y)$ mod $L$, this large diffeomorphism can be represented by the following shear operation:
\be
U_{\text{IR}}:\quad (x,y)\rightarrow (x+y, y).
\label{eq:U}
\ee
Then, for a mixed-state topological order $\rho_a$ in the anyon sector $a$, $U$ is a strong symmetry of $\rho_a$ with $U\rho=\theta_a\rho$. In the context of pure state topological phases, there have been proposals using such modular shear to extract chiral central charge, for example through momentum polarization \cite{Tu_2013}. For mixed states, one can start from the maximally mixed topological sector, 
\(
\rho = \sum_a \frac{d_a^2}{\mathcal{D}^2} \rho_a.
\)
Then the expectation value of modular shear $U$ naturally gives the chiral central charge via \eqref{eq:c},
\be
c_- = \frac{4}{\pi} \arg \big( \mathcal{D}\ \tr [ U \rho] \big).
\label{eq:c_U}
\ee
In the continuum, modular shear naturally preserves the space of closed anyon loops and simply deforms them. On the lattice, however, naïve application of \eqref{eq:U} can map closed loop to open strings which will be annihilated by $\rho$. Therefore, depending on the microscopic details, the explicit form of microscopic $U=U_{\text{UV}}$ needs to be modified and the modification can vary even within the same phase. For example, for both the  $\Z_N$ toric code defined on a square lattice and all the topological phases (chiral or nonchiral, twisted or untwisted) obtainable from condensing, decohering, or disordering the $\Z_N$ toric code, one can show that the natural definition of $U$ at fixed point is the generalization of the CNOT gate in qudits. This is discussed in Appendix \ref{app:shear}. For the decoherence channel which interpolates between the $\Z_3$ toric code and the chiral $\Z_3^{(1)}$ topological phase, we further present exact calculations of $\tr[U \rho]$ at arbitrary decoherence strength: at the zero-decoherence fixed point, we obtain $c_-=0$, while at the maximal-decoherence fixed point, $c_-=2$, consistent with the mathematical expectations. However, the calculation fails to capture the correct chirality structure away from fixed points, which is not surprising as \eqref{eq:c_U} is based on an expectation value linear in the density matrix, and any linear observable is not believed to capture the correct phase boundary. For \eqref{eq:c_U} to work at a general point in the phase, one needs to insert the dressed $\tilde{U}$ operator related to the dressed topological symmetries, which is in general hard to find.

\subsection{Boundary theory}
\label{subsec:bdry}

In pure states, chiral topological phases are associated with ungappable boundaries  \cite{ALVAREZGAUME1984269,hellerman2021,Fan_2022}. This intuition breaks down in mixed states. Below we will first illustrate this by constructing examples of short-range correlated boundaries of chiral mixed-state  topological phases. Then we examine the subtleties of the finite-depth two-way channel connecting the chiral $E_8$ mixed state to trivial product state. Based on this, we comment on the robustness of the gapless edge mode of $E_8$ state under mixed-state phase equivalence.

\subsubsection{Example for short-range correlated chiral boundary}

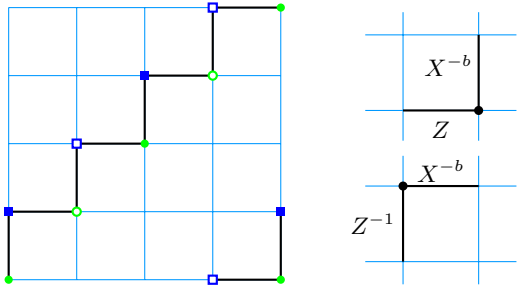
\begin{figure}[htbp]
\centering
\begin{tikzpicture}[scale=0.9]
\tikzset{grid line/.style={thin, color=blue!40!cyan}}
\draw[grid line] (0,0) grid (4,4);
\draw [black,thick] (0,0)--(0,1)--(1,1)--(1,2)--(2,2)--(2,3)--(3,3)--(3,4)--(4,4);
\draw [black, thick] (3,0)--(4,0)--(4,1);
\fill[color=blue] (-0.06,1-0.06) rectangle (0.06, 1+0.06);
\fill[color=blue] (2-0.06,3-0.06) rectangle (2+0.06, 3+0.06);
\fill[draw=blue,thick,fill=white] (3-0.06,0-0.06) rectangle (3+0.06, 0+0.06);
\fill[color=blue] (4-0.06,1-0.06) rectangle (4+0.06, 1+0.06);
\fill[draw=blue,thick,fill=white] (1-0.06,2-0.06) rectangle (1+0.06, 2+0.06);
\fill[draw=blue,thick,fill=white] (3-0.06,4-0.06) rectangle (3+0.06, 4+0.06);
\fill[color=green] (0,0) circle (0.06);
\fill[draw=green,fill=white,thick] (1,1) circle (0.06);
\fill[color=green] (2,2) circle (0.06);
\fill[draw=green,fill=white,thick] (3,3) circle (0.06);
\fill[color=green] (4,4) circle (0.06);
\fill[color=green] (4,0) circle (0.06);
\end{tikzpicture}
\qquad 
\begin{tikzpicture}[scale=1]
\tikzset{grid line/.style={thin, color=blue!40!cyan}}
\draw[grid line] (-0.5,-0.4) grid (1.5,1.3);
    \draw[thick, black] (1,0) -- (0,0); 
    \draw[thick, black] (1,0) -- (1,1);
    \fill[color=black] (1,0) circle (0.06);  
    \node at (0.5, -0.25) {$Z$};   
    \node at (0.6, 0.6) {$X^{-b}$};  
\tikzset{grid line/.style={thin, color=blue!40!cyan}}
\draw[grid line] (-0.5,-0.3-2.) grid (1.5,1.5-2.1); 
\draw[thick, black] (0,1-2) -- (0,0-2); 
    \draw[thick, black] (0,1-2) -- (1,1-2);  
    \node at (-0.4, -0.5-1) {$Z^{-1}$};  
    \node at (0.5, 1.2-2) {$X^{-b}$}; 
    \fill[color=black] (0,1-2) circle (0.06); 
\end{tikzpicture}
\caption{
Left: Support of the chiral $\Z_N$ topological phase on a cylinder. Start with $L$ by $L$ torus with even $L$, with the upper and lower sides identified, and left and right identified. Then cut the torus along the black thick staircase-shaped line to create a cylinder with two separate boundaries, which we will refer to as top and bottom boundaries respectively. See main text for explanation of the colored vertices. 
Right: Local boomerang-shaped  operators which generate the error channel. They are defined both in the bulk and on all vertices along the boundaries. 
}
\label{fig:bdry}
\end{figure}

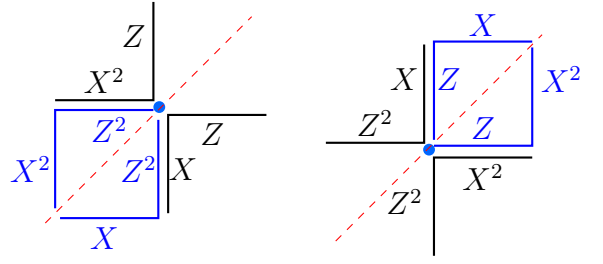
\begin{figure}[htbp]
\centering
\begin{tikzpicture}[scale=1.3]
\draw[thick,blue] (-1,-1)--(-1,0)--(0,0);
\draw[thick,blue] (0.05,-0.1)--(0.05,-0.1-1)--(0.05-1,-0.1-1);
\draw[thick,black] (-1,0+0.1)--(0,0+0.1)--(0,1+0.1);
\fill[color=blue!60!cyan] (0.06,0.03) circle (0.06);
\draw[thick,black] (0+0.15,-1-0.05)--(0+0.15,0-0.05)--(1.15,0-0.05);
\node[font=\large] at (-0.2,0.75) {$Z$};
\node[font=\large] at (-0.5, 0.3) {$X^2$};
\node[font=\large] at (0.6, -0.25) {$Z$};
\node[font=\large] at (0+0.3,-0.6) {$X$};
\node[font=\large] at (0.55-1,-0.2) {\textcolor{blue}{$Z^2$}};
\node[font=\large] at (-0.25-1,0.4-1) {\textcolor{blue}{$X^2$}};
\node[font=\large] at (-0.5,-1.3) {\textcolor{blue}{$X$}};
\node[font=\large] at (-0.15, -0.6) {\textcolor{blue}{$Z^2$}};
\draw[red, dashed] (-1.1,-1.1-0.05)--(1,1-0.05);
\end{tikzpicture}
\qquad
\begin{tikzpicture}[scale=1.3]
\draw[thick,blue] (0.1,0+0.13)--(0.1,1.13)--(1.1,1.13);
\draw[thick,black] (-1,0+0.1)--(0,0+0.1)--(0,1+0.1);
\fill[color=blue!60!cyan] (0.05,0.03) circle (0.06);
\draw[thick,black] (0+0.1,-1-0.05)--(0+0.1,0-0.05)--(1.1,0-0.05);
\draw[thick,blue] (1+0.1,1+0.07)--(1+0.1,0.07)--(0.1,0.07);
\node[font=\large] at (-0.2,0.75) {$X$};
\node[font=\large] at (-0.5, 0.3) {$Z^2$};
\node[font=\large] at (0.6, -0.25) {$X^2$};
\node[font=\large] at (0.6,1.3) {\textcolor{blue}{$X$}};
\node[font=\large] at (0.6,0.25) {\textcolor{blue}{$Z$}};
\node[font=\large] at (0+0.25,0.75) {\textcolor{blue}{$Z$}};
\node[font=\large] at (0-0.2,-0.5) {$Z^2$};
\node[font=\large] at (1.4, 0.75) {\textcolor{blue}{$X^2$}};
\draw[red,dashed] (-0.9,-0.9)--(1.2,1.2);
\end{tikzpicture}
\caption{The $O_v$ and $O_v'$ operators can be decomposed into boomerang-shaped open strings. These string operators don't all mutually commute, and whenever a convention fixing is needed, the blue strings are chosen to be applied before the black strings. Left: above the red dashed line, the two boomerangs comprise $O_v|_T$; below the red dashed line, the two boomerangs comprise $O_v|_B$. Right: above and below the red dashed line are the $O_v'|_T$ and $O_v'|_B$.}
\label{fig:decompose_O}
\end{figure}

Boundary theory of topological phases can be dependent on the shape, i.e. microscopic details of the boundary geometry. For our purpose, it is only necessary to show that there exists one boundary of an intrinsically mixed-state chiral topological phase, which  has only short-range correlations. Below we present a simple choice of a square lattice cylindrical system on the left panel of fig. \ref{fig:bdry}.
We will focus on the mixed-state topological stabilizer state with the only bulk local stabilizers being $O_v$ as defined in fig. \ref{fig:def_O}. In other words, the $O_v'$ sector is fully decohered out via error channels whose Kraus operators don't commute with $O_v'$ but commute with $O_v$. These Kraus operators are generated by the boomerang-shaped strings shown on the right panel of fig. \ref{fig:bdry}. We further fix a particular choice of $b$ such that the remaining mixed state is chiral, for instance, $b=1$ when $N=3$, or $b=2$ when $N=5$. To fully determine the mixed state, we specify all the stabilizers on the boundaries. Near the top and bottom boundaries along the black thick staircase in fig. \ref{fig:bdry}, the bulk operators need to be truncated. These truncated operators $O_v|_T$, $O_v'|_T$ near the top boundary, and $O_v|_B$, $O_v'|_B$ near the bottom boundary are defined in fig. \ref{fig:decompose_O}. 
The full density matrix for this chiral mixed state can now be written
\be
\rho_{\text{tot},\infty} \propto \prod_{v\in \text{bulk}} P_{v} \prod_{v_T\in \{\textcolor{green}{\bullet}\} } P^T_{v_T} \prod_{v_B \in \{\textcolor{blue}{\blacksquare}\}} P^B_{v_B},
\label{eq:truncated}
\ee
where $P_v=(\sum_{a=1}^N O_v^a)/N$ projects into the $O_v=1$ subspace in the bulk, $P^T_{v_T}$ projects into the $O_v|_T=1$ subspace, and  $P^B_{v_B}$ projects into $O_v|_B=1$. Similarly, one can even define the pre-decohered density matrix $\rho_{\text{tot},0}\propto \rho_{\text{tot},\infty}  \prod_{v\in \text{bulk}} P_{v}' \prod_{v_T\in \{{\color{green}\circ} \} } P^{\prime T}_{v_T} \prod_{v_B \in \{\textcolor{blue}{\square}\}} P^{\prime B}_{v_B}$. 
All the stabilizers in $\rho_{\text{tot},\infty}$ and $\rho_{\text{tot},0}$ mutually commute each other. Additionally, stabilizers of $\rho_{\text{tot},\infty}$ further commute with the local Kraus operators and can thus survive under decoherence. The bulk $O_v'$ and the truncated $O_v'|_{T,B}$ operators and don't commute with the Kraus operators as expected. 

One can check the counting is consistent: the total number of qudits is $2N_v+2\sqrt{N_v}$, where the second term $2\sqrt{N_v}$ comes from the boundary links and $N_v$ is the total number of vertices. As for the constraints, there are $N_v$ of full $O_v$ and $N_v$ full $O_v'$ operators. Since each type of truncated operators are only defined on every other vertices along the boundary, we have $\sqrt{N_v}/2$ for each of the following sets: $O_v|_T$, $O_v|_B$, $O_v'|_T$, $O_v'|_B$. They sum up to be $2N_v+2\sqrt{N_v}$ as expected. These are not all independent due to logical operators on the cylinder, but it is irrelevant for our consistency check. For $\rho_{\text{tot},\infty}$, there are only half, $N_v+\sqrt{N_v}$ constraints, consistent with the fact that the primed sector is in the maximally mixed state. 
Since the imposed stabilizers in \eqref{eq:truncated} all mutually commute, there is no algebraic decay of correlators and the boundary theory is not expected to be long-range correlated. Furthermore, since quantized thermal transport is closely tied to long-range correlated edge states based on prior experience in pure states \cite{Kitaev_2006,PhysRevB.25.2185, PhysRevB.55.15832}, we therefore expect that intrinsically mixed-state chiral topological phases do not have quantized thermal Hall conductivity.

\subsubsection{The $E_8$ state}
\label{subsec:E8}

The \(E_8\) phase hosts no nontrivial anyons and has chiral central charge \(c_-=8\) \cite{Kitaev_2006}. Therefore, it serves as a minimal example of chiral topological phase. Consider the pure-state \(E_8\) density matrix \(\rho_{E_8}=|\psi_{E_8}\rangle\langle\psi_{E_8}|\).  
The simplest local quantum channel to trivialize this mixed state is given by
\be
\E=\otimes_i \mathcal E_i,
\quad
\E_i[\rho_i]=\tr_i(\rho_i)\,\sigma_i
\ee
where \(i\) labels a local region, \(\rho_i\) is  supported on region \(i\), and \(\sigma_i\) is a fixed local density matrix, which WLOG can be chosen as \(|0\rangle\langle 0|\). Applied to the \(E_8\) density matrix, this trace-and-prepare channel gives
$
\mathcal E(\rho_{E_8})=\otimes_i \sigma_i ,
$
a trivial product state. 

Next we construct the reverse channel. Since the $E_8$ pure state is invertible \cite{Kitaev_2006,Plamadeala_2013,kong2014braidedfusioncategoriesgravitational}, its stacking with the time reversal partner \(E_8\oplus \overline E_8\) is trivial. Then there exists a \textit{quasi}-local unitary $\mathcal{C}$, such that starting from trivial product states $|0\rangle$ in both the system $S$ and the environment $E$, there is 
\(
\mathcal{C}|0\rangle_S|0\rangle_E
=
|E_8\rangle_S|\overline{E}_8\rangle_E.
\)
Therefore the channel
\be
\tilde{\E} (|0\rangle_S \langle 0|)
=
\operatorname{Tr}_E
\left[
\mathcal{C}
\left(
|0\rangle_S \langle 0| \otimes |0\rangle_E\langle 0|
\right)
\mathcal{C}^\dagger
\right]
\ee
prepares the pure \(E_8\) density matrix on \(S\). If the phase equivalence between two mixed states is defined through two-way connection via \textit{quasi-local} channels as in \cite{PRXQuantum.6.010315}, then the channels described above indicate that the the mixed-state $E_8$ is in the same trivial phase as the product state $|0\rangle \langle 0|$. In this case, since the product state can have short-range correlated boundaries, the gapless boundary of pure-state $E_8$ is not a robust feature throughout the same mixed-state phase. However, if mixed-state phase equivalence is defined instead through two-way connection via strictly \textit{local} finite depth channels, then the above reverse channel does not qualify and the $E_8$ state is then not equivalent to trivial product phase.

\subsection{Modular commutators}
\label{subsec:mod_com}

In pure state chiral topological phases, it is expected that the bulk entanglement spectrum one-to-one corresponds to the boundary energy spectrum \cite{Li_2008, PhysRevLett.108.196402}, and both are gapless. We have seen in the previous section \ref{subsec:bdry} that the boundary spectrum is no longer ``gapless'', or long-range correlated for mixed states, therefore, the bulk-boundary correspondence, if any, will more resemble that in pure-state nonchiral topological phases instead of the pure-state chiral topological phases \cite{Ho_2015,PhysRevB.99.205137}. We thus expect that the reduced density matrix of an intrinsically mixed-state chiral topological phase can have a boundary realization
with only short-range correlations. Below we will first show this explicitly in an example, and then use this result to compute the modular commutator -- the latter was proposed to detect chirality in pure state topological phases \cite{Kim_2022_0,Fan_2022,Kim_2022,vardhan2025}. 

Start from our old friend of the chiral $\Z_3^{(1)}$ topological phase defined on a torus, $\rho = 3^{-N_v-1} \prod_v P_v'$, where again $P_v$ projects to $O_v'=1$. One can partition the torus into two cylinders and trace out one of them. 
It is straightforward to check that the reduced density matrix $\rho_A$ is simply the product of all $O_v'$ projectors which fully live inside $A$. 
The same result holds if subsystem $A$ is not a cylinder but topologically trivial. 

For a tri-partite system $A, B, C$, the modular commutator is defined as \cite{Kim_2022_0} $\ii \tr \big( \rho_{ABC} [ K_{AB} (\rho), K_{BC} (\rho)]\big)$, where $K_X(\rho)=-\ln \rho_X$ is the modular Hamiltonian of the reduced density matrix $\rho_X$. For our example, since the reduced density matrices are all projectors of stabilizers $O_v'$ in various regions, using the fact that these operators mutually commute, we directly conclude that the modular commutator and its nested generalizations \cite{vardhan2025chiralitymagicquantumcorrelations} vanish. 

While the calculation is done for the chiral $\Z_3$ case, it's easily generalizable for chiral $\Z_N$ cases and we expect the same result both for these general examples and for general intrinsically chiral mixed-state topological phases. Therefore, the modular commutator is no longer a good measure for mixed-state topological phases - intuitively, this is because the modular flow in one subregion no longer affects the entanglement entropy in another region \cite{Fan_2022}, even if they have nontrivial overlap.

\section{Discussions}
\label{sec:discussion}

This work provides a comprehensive overview of various chirality measures for intrinsically mixed-state topological phases, clarifying their domains of applicability and limitations. By contrasting these measures with their pure-state counterparts, we highlight the exotic and intrinsically mixed-state features of topological systems. In particular, we propose two preferred relative entropy based measures that can unambiguously diagnose chirality in decohered topological phases. Although these measures are formulated for discrete error channels, we expect that their generalization to continuous-time evolution governed by Lindbladian dynamics will be equally effective. The limitation of the proposed measures  lies in the fact that they rely on prior knowledge of the error-free topological phase before decoherence. This assumption is natural in the context of fault-tolerant quantum computation, where one always has a target state or target phase in mind. However, for a general density matrix given outside this setting, we currently do not have a satisfactory measure that can unambiguously identify chirality without such prior information, especially beyond fixed points. A full understanding of this more ambitious problem is left for future work.

\textit{Note Added.} Near the completion of the draft we became aware of a related work which appeared on arXiv one week before the current work \cite{ellison2026manybodychiralitytopologicalstabilizer}. It views chirality in topological stabilizer states as an obstruction to transforming a state into its complex conjugate through finite-depth local operations. This view can be applied to mixed states and is in spirit aligned with the root idea in our section \ref{sec:OP}. However, our work is less proof-oriented and focuses instead on computable quantum measures for chirality in noisy topological matter that are valid beyond stabilizer states. 

\begin{acknowledgements}
We appreciate insightful discussions with Yimu Bao, Zhen Bi, Fiona Burnell, Meng Cheng, Ruihua Fan, Tarun Grover, Yingfei Gu, Chao-Ming Jian, Yuhan Liu, Abhinav Prem, Ramanjit Sohal, Kai Sun, Chong Wang, Sagar Vijay, Zhenghan Wang, Cenke Xu, and Carolyn Zhang. Zhu-Xi is particularly thankful to Tyler Ellison for pointing out an error in the draft. 
This research was supported in part by grant NSF PHY-2309135 to the Kavli Institute for Theoretical Physics. The manuscript is completed in part at the Simons Center for Geometry and Physics, Stony Brook University, and the Nordita Institute for Theoretical Physics.
\end{acknowledgements}

\appendix

\section{Details of the $\Z_N$ model}
\label{app:setup}

Here we argue that the unconventional definition of the $\Z_N$ toric code Hamiltonian using terms in fig.\  \ref{fig:def_O} is equivalent to the conventional definition using star $A_v$ and plaquette $B_p$ operators, as long as $\gcd (b,N)=1$. On a torus, the sets $\{O_v\}$ and $\{O_v'\}$ each satisfy a single global constraint,
$
\prod_v O_v = 1=\prod_v O_v',
$
but are otherwise independent. Thus the stabilizer group generated by $\{O_v, O_v'\}$ has rank $2N-2$. Recall the conventional $\Z_N$ toric code Hamiltonian written using the star and plaquette operators also gives the stabilizer group of the same rank. Moreover, since the $O_v$ and $O_v'$ operators are products of these star and plaquette operators, if the latter are stabilized, then the former must also be stabilized. 
Thus, the conventional $\Z_N$ toric-code ground-state subspace is contained in the ground-state subspace of the Hamiltonian~\eqref{eq:Ham}. Since the two subspaces have the same dimension, they must coincide. Therefore, indeed the Hamiltonian~\eqref{eq:Ham} realizes the same nonchiral $\mathbb{Z}_N$ toric code phase. 

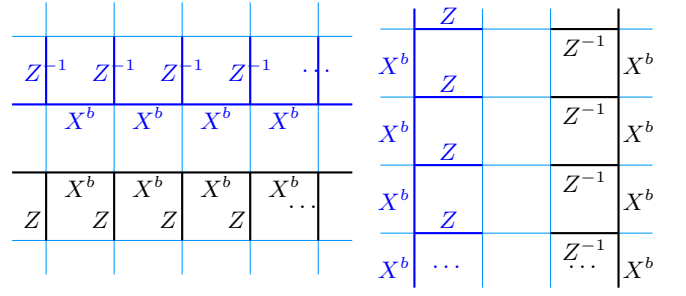
\begin{figure}[htbp]
\centering
\begin{tikzpicture}[scale=0.9]
    \tikzset{grid line/.style={thin, color=blue!40!cyan}}
    \draw[grid line] (-2.5,-1.5) grid (2.5,2.5);
    \draw[thick, blue] (-2.5, 1) -- (2.5, 1);
    \draw[thick, blue] (-1, 2) -- (-1, 1.0);
    \draw[thick, blue] (-2, 2) -- (-2, 1.0);
    \draw[thick, blue] (0, 2) -- (0, 1.0);
    \draw[thick, blue] (1, 2) -- (1, 1.0);
    \draw[thick, blue] (2, 2) -- (2, 1.0);
    \node at (-1.5, 0.8) {\textcolor{blue}{$X^b$}};   
    \node at (-0.5, 0.8) {\textcolor{blue}{$X^b$}};   
    \node at (0.5, 0.8) {\textcolor{blue}{$X^b$}};   
    \node at (-2, 1.5) {\textcolor{blue}{$Z^{-1}$}}; 
    \node at (-1.0, 1.5) {\textcolor{blue}{$Z^{-1}$}}; 
    \node at (0.0, 1.5) {\textcolor{blue}{$Z^{-1}$}};
    \node at (1.0, 1.5) {\textcolor{blue}{$Z^{-1}$}};
    \node at (2, 1.5) {\textcolor{blue}{$\dots$}};
    \node at (1.5, 0.8) {\textcolor{blue}{$X^b$}};
    \draw[thick] (-2.5, 0) -- (2.5, 0);
    \draw[thick] (-1, 0) -- (-1, -1);
    \draw[thick] (-2, 0) -- (-2, -1);
    \draw[thick] (0, 0) -- (0, -1);
    \draw[thick] (1, 0) -- (1, -1);
    \draw[thick] (2, 0) -- (2, -1);
    \node at (-1.5, -0.2) {{$X^b$}};   
    \node at (-0.5, -0.2) {{$X^b$}};   
    \node at (0.5, -0.2) {{$X^b$}};   
    \node at (1.5, -0.2) {{$X^b$}};
    \node at (-2.2, -0.7) {{$Z$}}; 
    \node at (-1.2, -0.7) {{$Z$}}; 
    \node at (-0.2, -0.7) {{$Z$}};
    \node at (0.8, -0.7) {{$Z$}};
    \node at (1.8, -0.5) {{$\dots$}};
    \draw[white] (0,-1.7)--(1,-1.7);
\end{tikzpicture}\ \
\begin{tikzpicture}[scale=0.9]
    \tikzset{grid line/.style={thin, color=blue!40!cyan}}
    \draw[grid line] (-1.5,-1.8) grid (2.5,2.3);
    \draw[thick] (2, -1.8) -- (2, 2.3);
    \draw[thick] (2, 2) -- (1.0, 2);
    \draw[thick] (2, -1) -- (1.0, -1);
    \draw[thick] (2, 1.0) -- (1.0, 1.0);
    \draw[thick] (2, 0.0) -- (1.0, 0.0);
    \node at (2+0.3, -1.5) {{$X^b$}};   
    \node at (2+0.3, -0.5) {{$X^b$}};   
    \node at (2+0.3, 0.5) {{$X^b$}};   
    \node at (2+0.3, 1.5) {{$X^b$}};  
    \node at (1.5, -1.25) {{$Z^{-1}$}};   
    \node at (1.5, 1-0.25) {{$Z^{-1}$}};    
    \node at (1.5, 2-0.25) {{$Z^{-1}$}};    
    \node at (1.5, -1.5) {{$\dots$}};
    \node at (1.5, -0.25) {{$Z^{-1}$}};
    \draw[thick,blue] (-1, -1.8) -- (-1, 2.3);
    \draw[thick,blue] (0, 2) -- (-1.0, 2);
    \draw[thick,blue] (0, -1) -- (-1.0, -1);
    \draw[thick,blue] (0, 1.0) -- (-1.0, 1.0);
    \draw[thick,blue] (0, 0.0) -- (-1.0, 0.0);
    \node[thick,blue] at (-1.3, -1.5) {\textcolor{blue}{$X^b$}};   
    \node at (-1.3, -0.5) {\textcolor{blue}{$X^b$}};   
    \node at (-1.3, 0.5) {\textcolor{blue}{$X^b$}};   
    \node at (-1.3, 1.5) {\textcolor{blue}{$X^b$}};  
    \node at (-0.5, -0.8) {\textcolor{blue}{$Z$}};   
    \node at (-0.5, 1.2) {\textcolor{blue}{$Z$}};    
    \node at (-0.5, 2.2) {\textcolor{blue}{$Z$}};    
    \node at (-0.5, -1.5) {\textcolor{blue}{$\dots$}};
    \node at (-0.5, 0.2) {\textcolor{blue}{$Z$}};  
\end{tikzpicture}
\caption{The logical operators defined on non-contractible loops of the torus. Left top: $W_x$ runs around the torus meridian and can be moved vertically by applying product of $O_v$. Left bottom: $W_x'$ an be moved vertically by applying product of $O_v'$. Right: The blue operator $W_y$ runs along the torus longitude and can be moved horizontally by applying product of $O_v$. The black operator $W_y'$ can be moved horizontally by applying product of $O_v'.$}
\label{fig:def_W}
\end{figure}

Now we define the logical operators of this model for and focus on the odd $N$ case -- it is known that the modular tensor category corresponding to the quantum double of $\Z_N$ with odd $N$ can be factorized into two modular tensor categories, and in certain cases, these two factors can have opposite chiralities. On a torus, there are $N^2$ superselection sectors in the ground state subspace, which can be distinguished by the eigenvalues of non-contractible loop operators $W_x, W_x'$ winding around the longitude of the torus. These operators are defined in fig.\ \ref{fig:def_W} and satisfy $[W_i, W_j']=0$ and \footnote{If $N$ is even, then these operators need to be modified to recover the full symmetry algebra of $\Z_N$ topological phase.}
\be
W_x W_y=\omega^{-2b} W_y W_x,\quad  W_x'W_y'=\omega^{2b} W_y'W_x'. 
\label{eq:algebra}
\ee

\section{Mapping relative entropy to the Potts model}
\label{app:Potts}

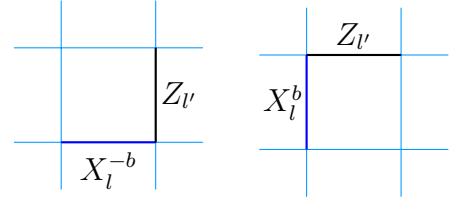
\begin{figure}[htbp]
\centering
\begin{tikzpicture}[scale=1.25]
\tikzset{grid line/.style={thin, color=blue!40!cyan}}
\draw[grid line] (-0.5,-0.5) grid (1.5,1.5);
    \draw[thick, blue] (-1+2,0) -- (-2+2,0); 
    \draw[thick, black] (-1+2,0) -- (-1+2,1); 
    \coordinate (v1) at (-1+2,0);
    \node[font=\large] at (-1.5+2, -0.3) {$X_{l}^{-b}$};   
    \node[font=\large] at (-1+2.25, 0.5) {$Z_{l'}$};  
\end{tikzpicture}\qquad
\begin{tikzpicture}[scale=1.25]
\tikzset{grid line/.style={thin, color=blue!40!cyan}}
\draw[grid line] (-1.5,-0.5) grid (0.5,1.5); 
    \draw[thick, blue] (1-2,0+1) -- (1-2,-1+1); 
    \draw[thick, black] (1-2,0+1) -- (2-2,0+1);  
    \coordinate (v2) at (1-2,0+1);
    \node[font=\large] at (1-2.25, -0.5+1) {$X_{l}^b$};  
    \node[font=\large] at (1.5-2, 0+1.2) {$Z_{l'}$};   
\end{tikzpicture}
\caption{Two elementary operators $S_l$ near the blue link $l$ which commute with $O_v'$ but not some of the $O_v$ operators. They don't mutually commute with each other. Notice that the midpoint of the link $l$ always differs from midpoint of $l'$ by a constant vector $(1/2,1/2)$ where the lattice constant is taken to be one. 
}
\label{fig:Kraus}
\end{figure}

Here we derive the statistical-mechanics model that controls the relative entropy diagnostic defined in Sec.~\ref{sec:rel_S}.  Start with the $\mathbb{Z}_N$ toric code,  we decohere the topological sector generated by {$em^{-b}$} but leave intact the sector generated by {$em^b$}. Incoherent noise is generated by local open string operators $S_l$ as shown in Fig.~\ref{fig:Kraus}, which originate from the boomerang decomposition of the $O_v$ operator in fig. \ref{fig:decompose_O}. They commute with the $O_v'$ stabilizers but not the $O_v$ stabilizers in fig. \ref{fig:def_O}. 
The specific error channel is 
\begin{equation}
    \begin{split}
        \mathcal{E}_\mu[\rho_0]&= (\otimes_l \mathcal{E}_{\mu,l}) [\rho_0],\\\mathcal{E}_{\mu,l} [\rho_0] &=\sum_{k=0}^{N-1} F_{k} (\mu) S_l^k \rho_0 S_l^{-k},
    \end{split}
    \label{eq:Z3channel}
\end{equation}
The $F$ functions characterize the noise strength and satisfy the normalisation $\sum_{k=0}^{N-1}F_k(\mu)=1$. Before decoherence,  $F_{0}(0)=1$ and $F_{k\neq 0}(0)=0$. As $\mu$ increases, the error probabilities $F_{k\neq 0} (\mu)$ grow monotonically with $\mu$. In the maximal-decoherence limit $\mu\rightarrow\infty$, all three occur with equal probability, $F_a(\infty)=1/N$. 
To evaluate the von Neumann relative entropy, we first introduce a Rényi-$n$ version of the relative entropy, defined as
\begin{equation}
    D^{(n)}(\rho_\mu||\rho_{a,\mu}) = \frac{1}{1-n}\log\frac{\tr \rho_\mu\rho_{a,\mu}^{n-1}}{\tr \rho_\mu^n}.
    \label{eq:RenyiRE}
\end{equation}
The von Neumann relative entropy is then obtained from the analytic continuation $n\rightarrow 1$. 
This replica formulation is useful because both the denominator and the numerator admit a natural loop-gas expansion: the parent $\Z_N$ toric-code state factorizes into equal weight superpositions of closed loop configurations of the $em^b$ and $em^{-b}$ anyons.

To be concrete, below we will focus on the $N=3$, $b=1$ case, and take the symmetric channel
\begin{equation}
    \begin{split}
        F_0(\mu) &= e^{-\mu}+\frac{1-e^{-\mu}}{3},\\
        F_1(\mu) &= F_2(\mu) = \frac{1-e^{-\mu}}{3}.
    \end{split}
\label{eq:F}
\end{equation}
Then the closed {$em^{-1}$} loop configurations map to the domain wall configurations of a coupled $(n-1)$-flavor three-state Potts model, as well as a dual random-bond three-state Potts model on the Nishimori line. This mapping is a direct extension of the methods developed in Ref.~\cite{PRXQuantum.5.020343} for decohered $\mathbb{Z}_2$ toric-code states.

\subsection{Mapping to replicated Potts model}

We derive the replicated statistical model directly from the Kraus
representation of the channel.  Denote a global error configuration by
$E=\{E_l\}$, with $E_l\in\mathbb Z_3$, and define
\begin{equation}
 K_{E}=\sqrt{P_\mu(E)}S(E),\quad
 P_\mu(E)=\prod_l F_{E_l}(\mu),
 \label{eq:global_kraus}
\end{equation}
where $S(E)=\prod_l^{\rightarrow}S_l^{E_l}$ with a fixed ordering
convention.  The decohered state can then be written as
$\rho_\mu=\sum_{E}K_{E}\rho_0 K_{E}^\dagger$. Expanding the denominator of Eq.~\eqref{eq:RenyiRE} gives
\begin{equation}
 \tr\rho_\mu^n
 =\sum_{\{E^{(r)}\}}
 \prod_{r=1}^nP_\mu(E^{(r)}) \tr\left[
 \prod_{r=1}^n
 \rho_0S^\dagger(E^{(r)})S(E^{(r+1)})
 \right],
 \label{eq:replica_kraus_expansion}
\end{equation}
where $E^{(n+1)}\equiv E^{(1)}$ and the product inside the trace is
ordered from left to right with increasing $r$.

The trace in Eq.~\eqref{eq:replica_kraus_expansion} is nonzero only when the
error strings in every pair of adjacent replicas have the same endpoints.
Equivalently,
\begin{equation}
  C^{(r)}=E^{(r+1)}-E^{(r)},
 \label{eq:adjacent_replica_loops}
\end{equation}
where $C^{(r)}$ are closed $\mathbb{Z}_3$ loops, and only $n-1$ of these loops are independent
\begin{equation}
    \sum_{r=1}^n C^{(r)}=0\pmod 3.
\end{equation}
Phases resulting from the ordering of the $S_l$ operators cancel around the replica cycle. Possible noncontractible loops change the boundary conditions of the statistical model but do not affect its bulk transition, and will be suppressed below.

Choose the error configuration in the first replica as a reference,
\begin{equation}
 E^{(r+1)}=E^{(1)}+g^{(r)},\qquad r=1,\ldots,n-1.
 \label{eq:reference_error}
\end{equation}
Each $g^{(r)}$ is a closed loop.  Up to a normalization factor
$\mathcal N$ that cancels in the relative entropy, the denominator becomes
\begin{equation}
 \tr\rho_\mu^n=
 \frac{1}{\mathcal N}
 \sum_{E^{(1)}}\sum_{\{g^{(r)}\}}
 P_\mu(E^{(1)})
 \prod_{r=1}^{n-1}P_\mu(E^{(1)}+g^{(r)}).
 \label{eq:replica_reference_sum}
\end{equation}
Here and below the sum over $g^{(r)}$ is restricted to closed-loop configurations. 

We next map the closed loops to domain walls of Potts spins. Introduce $n-1$ three-state variables $s_i^{(r)}\in\mathbb Z_3$ on the dual lattice, whose sites are located at the centers of plaquettes of the original lattice. For each link $l$ crossed by the dual link $\langle i,j\rangle$, set
\begin{equation}
 g_{l}^{(r)}=s_i^{(r)}-s_j^{(r)}\pmod 3
 \label{eq:loop_to_spin}
\end{equation}
This parametrization automatically imposes the closed-loop constraint. Additionally, we introduce random bond variables on the dual lattice
\begin{equation}
    \eta_{ij}=E_l^{(1)}\in\mathbb Z_3.
 \label{eq:reference_bond_variable}
\end{equation}
For the particle-hole symmetric channel, $F_1(\mu)=F_2(\mu)$.  The local weight in replica $r+1$ therefore distinguishes only whether the corresponding Potts bond is satisfied:
\begin{equation}
 F_{E_l^{(r+1)}}(\mu) =F_1(\mu)
 \exp\left[ K\delta(s_i^{(r)}-s_j^{(r)}+\eta_{ij})\right],
 \label{eq:RBP_local_weight}
\end{equation}
where the Nishimori coupling $K$ is defined by
\begin{equation}
 e^K=\frac{F_0(\mu)}{F_1(\mu)}
 =\frac{1+2e^{-\mu}}{1-e^{-\mu}}.
 \label{eq:e^K}
\end{equation}
The prefactor $F_1(\mu)$ is independent of the Potts spins and is absorbed into the normalization.  For fixed $E^{(1)}$, or equivalently fixed random bonds $\eta=\{\eta_{ij}\}$, the sum over each $g^{(r)}$ is the partition function of a random-bond three-state Potts model 
\begin{equation}
 Z_\eta
 =\frac{1}{\mathcal{N}}
 \sum_{\{s_i\}} e^{K\sum_{\langle ij\rangle}\delta(s_i^{(r)}-s_j^{(r)}+\eta_{ij})}.
 \label{eq:RBP_Z}
\end{equation}
Equation~\eqref{eq:replica_reference_sum} then becomes
\begin{equation}
 \tr\rho_\mu^n
 =\overline{Z_\eta^{\,n-1}},
 \label{eq:RBP_Zn}
\end{equation}
Using Eq.~\eqref{eq:e^K}, the single-bond distribution can be written as
\begin{equation}
 P(\eta_{ij})
 =p\,e^{K\delta_{\eta_{ij},0}},
 \qquad
 p=\frac{1}{e^K+2}.
 \label{eq:RBP_prob}
\end{equation}
Equations~\eqref{eq:RBP_Z} and \eqref{eq:RBP_prob} are precisely the random-bond three-state Potts model on the Nishimori line.  For fixed $\eta$, the $n-1$ Potts copies are decoupled, they are correlated only through their common bond realization.  At finite integer $n$, Eq.~\eqref{eq:RBP_Zn} is an annealed replicated disorder average, while the $n\rightarrow1$ limit gives the quenched average.

Moving on to the numerator of Eq.~\eqref{eq:RenyiRE}.  Let $L_{em^{-1}}(\ell)$ be an open string whose endpoints are the dual sites $i$ and $j$ and denote its $\mathbb{Z}_3$ link configuration by $L = \{L_l\}$. Take the first replica to be $\rho_\mu$ and the remaining $n-1$ replicas to
be $\rho_{em^{-1},\mu}$.  A nonzero trace requires
\begin{equation}
E^{(r+1)}=E^{(1)} + g^{(r)}+L,
\label{eq:numerator_matching_condition}
\end{equation}
The numerator is then
\begin{equation}
  \tr\rho_\mu\rho_{em^{-1},\mu}^{n-1} = \frac{1}{\mathcal N}
  \sum_{E^{(1)}}\sum_{\{g^{(r)}\}} P_\mu(E^{(1)})\cdot
  \prod_{r=1}^{n-1}
  P_\mu(E^{(1)}+g^{(r)} +L).
 \label{eq:numerator_reference_error_sum}
\end{equation}
For the same bond realization $\eta_{ij}=E_l^{(1)}$, define
\begin{equation}
 Z_\eta[L_{em^{-1}}]=
 \frac{1}{\mathcal{N}}
 \sum_{\{s_i\}}
 e^{K\sum_{\langle ij\rangle}\delta(s_i^{(r)}-s_j^{(r)}+\eta_{ij}+L_{ij})}.
 \label{eq:twisted_bond_weight}
\end{equation}
The numerator becomes
\begin{equation}
 \tr\rho_\mu\rho_{em^{-1},\mu}^{n-1}
 =\overline{\left(Z_\eta[L_{em^{-1}}]\right)^{\,n-1}}.
 \label{eq:twisted_potts_partition}
\end{equation}
With the parametrization in Eq.~\eqref{eq:numerator_matching_condition}, no additional braiding phase needs to be introduced.  For a fixed ordering convention, the error operator can be rearranged as
\begin{equation}
    S(E^{(r+1)}) = e^{i\phi_E^{r}} S(E^{(1)}) S(g^{(r)})L^\dagger_{em^{-1}}.
\end{equation}
where $e^{i\phi_E^{(r)}}$ accounts for the noncommutativity of the string operators.  These phases cancel between each error operator and its Hermitian conjugate in the cyclic replica trace, so no additional phase factor is required.

The bond shift in Eq.~\eqref{eq:twisted_bond_weight} can be understood as an open string insertion in the Potts representation. Combining  Eqs.~\eqref{eq:RBP_Zn} and \eqref{eq:twisted_potts_partition}, the R\'enyi relative entropy is
\begin{equation}
 D_n(\rho_\mu\Vert\rho_{em^{-1},\mu})
 =\frac{1}{1-n}\log
 \frac{\overline{\left(Z_\eta[L_{em^{-1}}]\right)^{n-1}}}
 {\overline{Z_\eta^{n-1}}}.
 \label{eq:RBP_Renyi_relative_entropy}
\end{equation}
Taking $n\rightarrow1$ in Eq.~\eqref{eq:RBP_Renyi_relative_entropy} gives
\begin{equation}
 D(\rho_\mu\Vert\rho_{em^{-1},\mu})
 =\overline{\log Z_\eta-
 \log Z_\eta[L_{em^{-1}}]}.
 \label{eq:RBP_defect_free_energy}
\end{equation}
Thus the relative entropy is the disorder-averaged excess free energy of the defect line. Since the Nishimori coupling $K$ decreases as the decoherence strength $\mu$ increases, weak decoherence corresponds to the ferromagnetic phase of the random-bond Potts model. In this phase, the defect line requires a $\mathbb Z_3$ domain wall connecting its endpoints. The domain wall has a nonzero tension and a minimum length proportional to $\ell$, giving $D(\rho_\mu\Vert\rho_{em^{-1},\mu})\propto\ell$. At strong decoherence, the Potts model is instead paramagnetic. Domain walls then proliferate and screen the defect, so its free-energy cost is localized near the two endpoints. Consequently, the relative entropy approaches a constant as $\ell\rightarrow\infty$.

The transition point is set by the Nishimori transition of the random-bond three-state Potts model, which occurs at~\cite{Jacobsen_2002}
\begin{equation}
p_c\simeq0.0785,\quad K_c
=
\log\left(\frac{1}{p_c}-2\right) \simeq2.3739.
\end{equation}
This corresponds to the critical decoherence strength
\begin{equation}
    \mu_c = \log\left(\frac{3}{e^{K_c}-1}+1 \right)\simeq 0.2682.
    \label{eq:mu_cRBP}
\end{equation}

For finite integer $n$, the transition can be analyzed using the
duality argument of Nishimori and Nemoto~\cite{Nishimori_2002}. The invariant-subspace condition is
\begin{equation}
 \sum_{q=0}^{2} F_q^n = 3^{-(n-1)/2}.
 \label{eq:finite_n_duality}
\end{equation}
For $n=2$, the disorder-averaged model contains a single Potts variable and reduces to the ordinary three-state Potts model.  Its bond weights form a one-parameter family that is closed under the duality, so Eq.~\eqref{eq:finite_n_duality} gives the exact critical point
\begin{equation}
 \mu_c^{(2)}
 =\frac{1}{2}\log(1+\sqrt{3})
 \simeq 0.5025.
 \label{eq:n2_transition}
\end{equation}
For $n=3$, the bond weights is also closed under duality, so Eq.~\eqref{eq:finite_n_duality} again gives the exact transition point
\begin{equation}
 \mu_c^{(3)}
 =\log\left(2\cos\frac{\pi}{9}\right)
 \simeq0.6309.
 \label{eq:n3_transition}
\end{equation}
For $n\geq4$, duality acts within a multidimensional parameter space, so Eq.~\eqref{eq:finite_n_duality} defines only an invariant subspace rather than a self-dual point. It therefore provides a conjecture for the transition, whose precise location must generally be determined numerically.

\subsection{Partical-hole asymmetric decoherence}

In this section, we generalize the previous mapping to a polarized channel
\begin{equation}
    \begin{split}
        F_0(\mu) &= e^{-\mu}+\frac{1-e^{-\mu}}{3},\\
        F_1(\mu) &= (1+x)\cdot\frac{1-e^{-\mu}}{3}\\
        F_2(\mu) &= (1-x)\cdot\frac{1-e^{-\mu}}{3},
    \end{split}
\label{eq:F_asy}
\end{equation}
with $-1\leq x\leq 1$. The two non-trivial error processes, corresponding to the creation of  $em^{-1}$ and $e^{-1}m$ anyon pairs, occur with unequal probabilities.

The replica matching conditions in Eqs.~\eqref{eq:adjacent_replica_loops} and \eqref{eq:reference_error}, as well as the domain-wall parametrization in Eq.~\eqref{eq:loop_to_spin}, are unchanged.  The asymmetric local weight can be written as
\begin{equation}
 F_q=A\exp\left[K\delta_{q,0}+h\nu(q)\right],
 \label{eq:asymmetric_local_weight}
\end{equation}
where
\begin{equation}
    \nu(q)
    =
    \begin{cases}
        0,  & q=0,\\
        1,  & q=1,\\
        -1, & q=2.
    \end{cases}
\end{equation}
and
\begin{equation}
    A=\sqrt{F_1F_2},\quad
    K=\log\frac{F_0}{\sqrt{F_1F_2}},\quad
    h=\frac{1}{2}\log\frac{F_1}{F_2}.
\label{eq:asymmetric_couplings}
\end{equation}
For the polarized channel \eqref{eq:F_asy}, 
\begin{equation}
    K=\log\left[ \frac{1+2e^{-\mu}} {(1-e^{-\mu})\sqrt{1-x^2}} \right], \,\, h=\frac{1}{2}\log \frac{1+x}{1-x}.
\end{equation}

For a fixed reference configuration $E^{(1)}$, define $\eta_{ij}=E_l^{(1)}$ and $q_{ij}=s_i-s_j+\eta_{ij}\pmod 3$.  The partition function in Eq.~\eqref{eq:RBP_Z} is then replaced by
\begin{equation}
 Z_\eta=
 \frac{1}{\mathcal N}\sum_{\{s_i\}}
 \exp\left\{\sum_{\langle ij\rangle}
 \left[K\delta_{q_{ij},0}+h\nu(q_{ij})\right]\right\},
 \label{eq:polarized_RBP_Z}
\end{equation}
with the bond distribution
\begin{equation}
 P(\eta_{ij}=q)=F_q
 =\frac{e^{K\delta_{q,0}+h\nu(q)}}{e^K+2\cosh h}.
 \label{eq:polarized_RBP_prob}
\end{equation}
Equations~\eqref{eq:polarized_RBP_Z} and \eqref{eq:polarized_RBP_prob} define a polarized random-bond three-state Potts model on a generalized Nishimori surface.  The replica expressions \eqref{eq:RBP_Renyi_relative_entropy} and \eqref{eq:RBP_defect_free_energy} remain valid with this partition function, so the relative entropy is again controlled by the excess free energy of the defect line.

The critical line of the polarized model is not presently known.  Applying the Nishimori--Nemoto conjecture~\cite{Nishimori_2002} in the $n\rightarrow1$ limit gives
\begin{equation}
 \sum_{q=0}^2F_q\log F_q=-\frac{1}{2}\log3.
 \label{eq:asymmetric_Nishimori_limit}
\end{equation}
For example, Eq.~\eqref{eq:asymmetric_Nishimori_limit} gives 
\begin{equation}
    \mu_c^{\mathrm{conj.}}\simeq0.2923
\end{equation}
at $x=0.5$. At finite $n$, the corresponding conjecture is Eq.~\eqref{eq:finite_n_duality}. For $n=2$, the model reduces to the ordinary three-state Potts model after disorder average. The duality condition is therefore exact and gives
\begin{equation}
    \mu_c^{(2)}\simeq0.5217,
\end{equation}
at $x=0.5$.  For $x\neq0$ and
$n\geq3$, the asymmetric theory is not self-dual, so Eq.~\eqref{eq:finite_n_duality} is only a conjecture and the transition should generally be determined numerically.

\section{Linear modular shear}
\label{app:shear}

\begin{figure}[htbp]
\centering
\begin{tikzpicture}[scale=0.6, 
    string/.style={thick},
    dot/.style={circle, fill=black, inner sep=0pt, minimum size=5pt},
    opencircle/.style={circle, draw=black, thick, inner sep=0pt, minimum size=5pt, fill=white}
]
    \node[opencircle] (tl_start) at (0, 0-0.5) {};
    \draw[string] (tl_start) -- (1.5, 0-0.5) node[midway, above, yshift=2pt] {$X$}; 
    \node at (2.5, 0-0.5) {$\longmapsto$};
    \node at (2.5, 0.4-0.5) {$U$};   
    \node[dot] (tl_end) at (4, 0-0.5) {};
    \draw[string] (tl_end) -- (5.5, 0-0.5) node[midway, above, yshift=2pt] {$X$};
    \node[opencircle] (bl_start) at (0, -3) {};
    \draw[string] (bl_start) -- (1.5, -3) node[midway, above, yshift=2pt] {$Z$};
\node at (2.5, -3) {$\longmapsto$};  \node[dot] (bl_end) at (4, -3+0.5) {};
    \draw[string] (bl_end) -- (5.5, -3+0.5) node[midway, above, yshift=2pt] {$Z$};
    \draw[string] (bl_end) -- (4, -4.5+0.5) node[midway, left, xshift=2pt] {$Z^{-1}$};
    \draw[dashed, darkgray] (6.5, 1) -- (6.5, -4.5);
    \draw[string] (8, 0.5) -- (8, -1) node[midway, left, xshift=-2pt] {$X$};
    \node[opencircle] (tr_start) at (8, -1) {};

    \node at (9., -0.25) {$\longmapsto$};
    
    \node[dot] (tr_end) at (11.5-1, -1) {};
    \draw[string] (11.5-1, -1+1.5) -- (11.5-1, -1) node[midway, left, xshift=0pt] {$X$}; 
    \draw[string] (11.5-1, -1+1.5) -- (12, -1+1.5) node[midway, below, xshift=2pt] {$X$}; 
    
    \draw[string] (8, -2.5) -- (8, -4) node[midway, left, xshift=-2pt] {$Z$};
    \node[opencircle] (br_start) at (8, -4) {};

    \node at (9, -3.75+0.5) {$\longmapsto$};
    
    \node[dot] (br_end) at (10.5, -4) {};
    \draw[string] (10.5, -2.5) -- (10.5, -4) node[midway, right, xshift=2pt] {$Z$};
\end{tikzpicture}
\caption{Definition of the action of modular shear $U$ on the local operators. Suppose the open circle has coordinates $(x,y)$, then the closed circle has coordinate $(x+y,y)$. This is the $\Z_N$ CNOT or CX gate  with vertical control and horizontal target qudits and therefore a 2-qudit unitary.}
\label{fig:lattice_shear}
\end{figure}
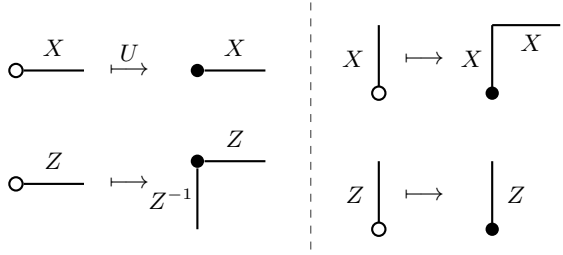

Here we compute the expectation values $\tr[U\rho]$ for the modular shear $U$ discussed in section \ref{subsec:shear} for decohered $\Z_N$ toric code, which could be either chiral or nonchiral. While in the continuum the modular shear unambiguously sends $(x,y)\mapsto (x+y,y)$, on the lattice the discrete cutoff requires more care. We define its action as in fig. \ref{fig:lattice_shear}, such that its action on $O_v$ and $O_v'$ simply shifts them in the $x$-direction by $y$ steps. 
With this definition, the shear acts on the long strings as
\be
\begin{split}
& U W_x U^{-1}= W_x,\quad U W_x' U^{-1} = W_x',\\
& U W_y' U^{-1}= \omega^{-b} W_x' W_y' \prod_{v\in \textcolor{blue}{R}} O_v', \\
& U W_y U^{-1}= \omega^{b} W_x W_y 
\big(
\prod_{v'\in \textcolor{green}{R'}} O_v' O_v^{-1} 
\big)
\big(\prod_{v\in \textcolor{blue}{R}} O_v^{-1}\big). 
\end{split}
\label{eq:U_on_W}
\ee
The blue region $\textcolor{blue}{R}$ corresponds to the lower right triangular region below the sheared diagonal operator, while the green region $\textcolor{green}{R'}$ corresponds to  every other vertex along the diagonal sheared operator. These last two equations are shown in fig. \ref{fig:shear1} and fig. \ref{fig:shear2}. 
\begin{figure}[htbp]
\centering
\begin{tikzpicture}[scale=1]
    \tikzset{grid line/.style={thin, color=blue!40!cyan}}
    \draw[grid line] (0,0) grid (6,6);
    \draw [black,thick] (0,0)--(0,1)--(1,1)--(1,2)--(2,2)--(2,3)--(3,3)--(3,4)--(4,4)--(4,5)--(5,5)--(5,6)--(6,6);
\fill[color=blue] (1,1) circle (0.06);
\fill[color=blue] (2,1) circle (0.06);
\fill[color=blue] (3,1) circle (0.06);
\fill[color=blue] (4,1) circle (0.06);
\fill[color=blue] (5,1) circle (0.06);
\fill[color=blue] (2,2) circle (0.06);
\fill[color=blue] (3,2) circle (0.06);
\fill[color=blue] (4,2) circle (0.06);
\fill[color=blue] (5,2) circle (0.06);
\fill[color=blue] (3,3) circle (0.06);
\fill[color=blue] (4,3) circle (0.06);
\fill[color=blue] (5,3) circle (0.06);
\fill[color=blue] (4,4) circle (0.06);
\fill[color=blue] (5,4) circle (0.06);
\fill[color=blue] (5,5) circle (0.06);
\node at (-0.4,0.4) {$ZX^b$};
\node at (1-0.4,1.4) {$ZX^b$};
\node at (2-0.4,2.4) {$ZX^b$};
\node at (3-0.4,3.4) {$ZX^b$};
\node at (4-0.4,4.4) {$ZX^b$};
\node at (5-0.4,5.4) {$ZX^b$};
\node at (0.5,0.75) {$Z^{-1}X^b$}; 
\node at (1.5,1.75) {$Z^{-1}X^b$}; 
\node at (2.5,2.75) {$Z^{-1}X^b$}; 
\node at (3.5,3.75) {$Z^{-1}X^b$}; 
\node at (4.5,4.75) {$Z^{-1}X^b$}; 
\node at (5.5,5.75) {$Z^{-1}X^b$}; 
\draw[thick,black] (5,0)--(6,0)--(6,1);
\end{tikzpicture}
\caption{Illustration of the second line in equation \eqref{eq:U_on_W}. Periodic boundary conditions are imposed on both directions. Black string is the sheared $U W_y' U^{-1}$ operator. Blue vertices are those belonging to region $\textcolor{blue}{R}$.}
\label{fig:shear1}
\end{figure}
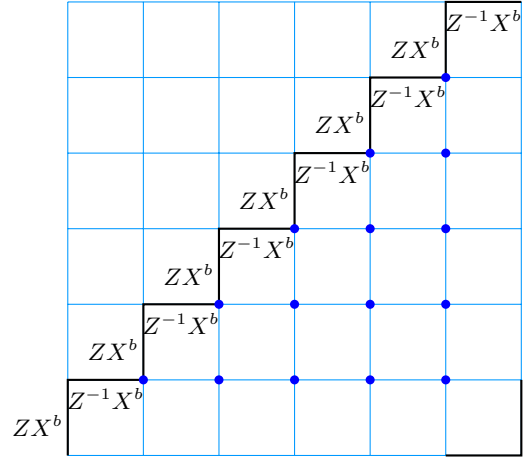
\begin{figure}[htbp]
\centering
\begin{tikzpicture}[scale=1]
    \tikzset{grid line/.style={thin, color=blue!40!cyan}}
    \draw[grid line] (0,0) grid (6,6);
    \draw [black,thick] (0,0)--(0,1)--(1,1)--(1,2)--(2,2)--(2,3)--(3,3)--(3,4)--(4,4)--(4,5)--(5,5)--(5,6)--(6,6);
    \draw [magenta,thick] (0,0)--(1,0)--(1,1)--(2,1)--(2,2)--(3,2)--(3,3)--(4,3)--(4,4)--(5,4)--(5,5)--(6,5)--(6,6);
    \draw[thick,magenta] (0,5)--(0,6)--(1,6);
\fill[color=green] (0,0) circle (0.06);
\fill[color=blue] (2,1) circle (0.06);
\fill[color=blue] (3,1) circle (0.06);
\fill[color=blue] (4,1) circle (0.06);
\fill[color=blue] (5,1) circle (0.06);
\fill[color=green] (2,2) circle (0.06);
\fill[color=blue] (3,2) circle (0.06);
\fill[color=blue] (4,2) circle (0.06);
\fill[color=blue] (5,2) circle (0.06);
\fill[color=blue] (4,3) circle (0.06);
\fill[color=blue] (5,3) circle (0.06);
\fill[color=green] (4,4) circle (0.06);
\fill[color=blue] (5,4) circle (0.06);
\fill[color=blue] (6,1) circle (0.06);
\fill[color=blue] (6,2) circle (0.06);
\fill[color=blue] (6,3) circle (0.06);
\fill[color=blue] (6,4) circle (0.06);
\fill[color=blue] (6,5) circle (0.06);
\fill[color=green] (6,6) circle (0.06);
\node at (-0.3,0.4) {$X^b$};
\node at (1-0.3,1.4) {$X^b$};
\node at (2-0.3,2.4) {$X^b$};
\node at (3-0.3,3.4) {$X^b$};
\node at (4-0.3,4.4) {$X^b$};
\node at (5-0.3,5.4) {$X^b$};
\node at (0.5,0.75) {$X^b$}; 
\node at (1.5,1.75) {$X^b$}; 
\node at (2.5,2.75) {$X^b$}; 
\node at (3.5,3.75) {$X^b$}; 
\node at (4.5,4.75) {$X^b$}; 
\node at (5.5,5.75) {$X^b$};
\node at (1-0.35,0.4) {\textcolor{magenta}{$Z^{-1}$}};
\node at (2-0.35,1.4) {\textcolor{magenta}{$Z^{-1}$}};
\node at (3-0.35,2.4) {\textcolor{magenta}{$Z^{-1}$}};
\node at (4-0.35,3.4) {\textcolor{magenta}{$Z^{-1}$}};
\node at (5-0.35,4.4) {\textcolor{magenta}{$Z^{-1}$}};
\node at (6-0.35,5.4) {\textcolor{magenta}{$Z^{-1}$}};
\node at (0.5,0.75-1) {\textcolor{magenta}{$Z$}}; 
\node at (1.5,0.75) {\textcolor{magenta}{$Z$}}; 
\node at (2.5,1.75) {\textcolor{magenta}{$Z$}}; 
\node at (3.5,2.75) {\textcolor{magenta}{$Z$}}; 
\node at (4.5,3.75) {\textcolor{magenta}{$Z$}}; 
\node at (5.5,4.75) {\textcolor{magenta}{$Z$}}; 
\draw[thick,black] (5,0)--(6,0)--(6,1);
\end{tikzpicture}
\caption{Illustration of the third line in equation \eqref{eq:U_on_W}. Black and magenta diagonal strings are the sheared $U W_y U^{-1}$. Blue vertices on the lower right triangular region live inside $\textcolor{blue}{R}$, and green vertices which live on every other vertex of the diagonal line are in region $\textcolor{green}{R'}$.}
\label{fig:shear2}
\end{figure}
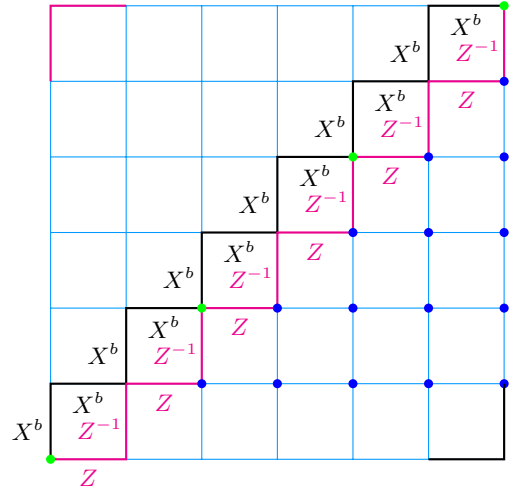

\subsection{At fixed points}

We are now ready to evaluate $\tr (\rho_0 U)$ where $\rho_0$ is stabilized by all $O_v$ and $O_v'$ operators, and is further maximally mixed in the $N^2$ logical sectors:
\be
\rho_0 = \frac{1}{N^2}\sum_{s,s'=0}^2 |\psi_{s,s'}\rangle \langle \psi_{s,s'}|,
\label{eq:pre-dec_1}
\ee
where $W_x|\psi_{s,s'}\rangle = \omega^s |\psi_{s,s'}\rangle$ and $W_x'|\psi_{s,s'}\rangle = \omega^{s'} |\psi_{s,s'}\rangle$. Using \eqref{eq:algebra}, it's straightforward to check that 
\be
\begin{split}
& W_x W_y |\psi_{s,s'}\rangle =\omega^{s-2b} W_y |\psi_{s,s'}\rangle,\\
& W_x' W_y' |\psi_{s,s'}\rangle =\omega^{s'+2b} W_y' |\psi_{s,s'}\rangle,
\end{split}
\label{eq:Wy_shift}
\ee
such that $W_y |\psi_{s,s'}\rangle \propto |\psi_{s-2b,s'}\rangle$ and $W_y' |\psi_{s,s'}\rangle \propto |\psi_{s,s'+2b}\rangle$. The proportionality phase factors do not matter.  Since $W_x, W_x'$ and Hamiltonian $H$ are invariant under the shear $U$, $U$ is diagonal in the $|\psi_{s,s'}\rangle$ basis; assume its eigenvalues are $\lambda_{s,s'}$. We now use equation \eqref{eq:U_on_W} to derive the relationships between them. First, in the ground state subspace, all the $O_v$ and $O_v$'s evaluate to one. So $U W_y |\psi_{s,s'}\rangle = \omega^{b} W_x W_y U |\psi_{s,s'}\rangle$, and $U W_y' |\psi_{s,s'}\rangle = \omega^{-b} W_x' W_y' U |\psi_{s,s'}\rangle$, such that 
\be
\lambda_{s-2b,s'}= \lambda_{s,s'} \omega^{s-b},\quad \lambda_{s,s'+2b}=\lambda_{s,s'}\omega^{s'+b}. 
\label{eq:iteration}
\ee
Therefore, there is only one independent eigenvalue. Combining \eqref{eq:pre-dec_1} and \eqref{eq:iteration}, for any choice of $b$, we can arrive at 
\be
\tr (U \rho_0)=\frac{1}{N^2}\sum_{s,s'}\lambda_{s,s'}=\frac{\lambda_{0,0}}{N}.
\ee
To pin down the value for $\lambda_{0,0}$, first notice that $U$ is a product of two-qudit unitaries (the CX gate) and therefore its eigenvalues must be norm one. The phase can be traced back to the fact that $|\psi_{0,0}\rangle$ includes the vector $|\bm{0}\rangle=|00\cdots 00\rangle$ in the computational basis, and it is natural to fix its coefficient to be real. Since CX leaves $|\bm{0}\rangle$ strictly invariant, it has to be that $\lambda_{0,0}=1.$ We finally arrive at
\be
c^- (0) = \frac{4}{\pi} \arg [\tr (\rho_0 U)]=0.
\label{eq:c0}
\ee
This is consistent with the expectation that $\Z_N$ toric code is nonchiral.

Next we examine $\tr[U \rho_{\infty}]$ at the other fixed point, with maximally decohered $\rho$. Since this value depends on the decoherence channel, to be concrete we focus on the case of $N=3$, and decohere out all anyons generated by {$em^{-1}$}. In other words, choosing $b=1$ in fig.\ \ref{fig:def_O} as the representation of the toric code, the Kraus operators do not commute with the $O_v$ terms, but do commute with the $O_v'$ terms. The surviving anyons are $\{1, em, e^{-1}m^{-1}\}$, which generates modular tensor category of rank $3$. Therefore, at maximal decoherence, this sector has no excitations, while the other sector $\{1,em^{-1},e^{-1}m\}$ becomes the infinite temperature state. The density matrix can be written in the following basis, 
\be
\rho_{\infty} = \frac{1}{3} \sum_{s'=0}^2 \rho_{s'},\quad U \rho_{s'} =\lambda_{0,s'} \rho_{s'}.  
\ee
The maximal mixture of different topological sectors is reminiscent of \eqref{eq:pre-dec_1}. Therefore, the chiral central charge reads
\be
c^-(\infty)=\frac{4}{\pi} \arg [1+2\omega]=2, 
\label{eq:c_infty}
\ee
consistent with the expectations from category theory.

\subsection{Wrong prediction of the transition point} 

Next we examine the $\tr[U \rho]$ calculation for general decoherence strength. We will focus on $N=3$, $b=1$ and use the channel in \eqref{eq:Z3channel} \eqref{eq:F} and fig. \ref{fig:Kraus}. Although the local boomerang operators $S$ themselves do not mutually commute, the local channels $\mathcal E_\ell$ do commute with one another. Therefore we can write
\begin{equation}
\mathcal E[\rho_0]
=
\sum_{\{a_l\}}
\Bigl(\prod_{j=1}^{M} F_{a_j}(\mu)\Bigr)
S_{\ell_{M}}^{a_{M}}\cdots S_{\ell_1}^{a_1}\,
\rho_0\,
S_{\ell_1}^{-a_1}\cdots S_{\ell_{M}}^{-a_{M}},
\label{eq:channel_long}
\end{equation}
where $M=2L^2$ is the total number of qutrits in the system.   
The global Kraus operators of the full channel can then be represented as 
\begin{equation}
\rho_{\mu} \equiv \E [\rho_0] = \sum_{\bm{a}} K_{\bm{a}}\rho_0 K_{\bm{a}}^{\dagger},\ \ K_{\mathbf a}(\mu)
=
\prod_\ell \sqrt{F_{a_\ell}(\mu) }
S_\ell^{a_\ell},
\label{eq:full-kraus}
\end{equation}
with any fixed ordering convention for the product over links. It's straightforward to check that
$\sum_{\bm{a}} K_{\bm{a}}^{\dagger}K_{\bm{a}}=I. 
$
Therefore, the expectation value of $U$ in $\rho_{\mu}$ is equal to the expecation value of $\mathcal{E}^{-1}[U]$ in $\rho_0,$
\be
\tr [U \rho_{\mu}] =\tr \big[\rho_0 \sum_{\bm{a}}  K_{\bm{a}}^{\dagger}U K_{\bm{a}} \big].
\ee

\begin{figure}[htbp]
\centering
\begin{tikzpicture}[scale=1.25]
\tikzset{grid line/.style={thin, color=blue!40!cyan}}
\draw[grid line] (-0.5,-0.5) grid (1.5,1.5);
    \draw[thick, black] (-1+2,0) -- (-2+2,0); 
    \draw[thick, black] (-1+2,0) -- (-1+2,1); 
    \coordinate (v1) at (-1+2,0);   
    \node[font=\large] at (-1.5+2, -0.3) {$X^{-1}$};   
    \node[font=\large] at (-1+2.25, 0.5) {$Z$};  
    \fill[color=blue] (0,0) circle (0.06);
    \node at (1-0.6,0.25) {\textcolor{blue}{$(x+y,y)$}};
\end{tikzpicture}\quad
\begin{tikzpicture}[scale=1.25]
\tikzset{grid line/.style={thin, color=blue!40!cyan}}
\draw[grid line] (-1.5,-0.5) grid (1.5,1.5); 
    \draw[thick, black] (0,1)--(-1,1)--(-1,0);
    \draw[thick,black] (1,1)--(0,1)--(0,0);
    \node[font=\large] at (1-2.2, -0.55+1) {$X$};  
    \node[font=\large] at (1-1.3, -0.55+1) {$Z^{-1}$};  
    \node[font=\large] at (1.5-2, 0+1.2) {$X$};
    \node[font=\large] at (1.5-1, 0+1.2) {$Z$};   
    \fill[color=blue] (-1,0) circle (0.06);
    \node at (-0.45,0.75-1) {\textcolor{blue}{$(x+y,y)$}};
\end{tikzpicture}
\caption{Definition of the sheared $S'_{l'}$ operator in \eqref{eq:S'}. If originally, before shear, the coordinates of the boomerang endpoint are $(x,y)$, then post-shear  they are shifted to $(x+y,y)$. }
\label{fig:sheared_S}
\end{figure}
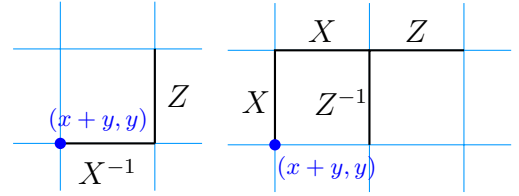

Let's start by examining how a single $S_l$ acts on $U.$ If $l$ is a horizontal link, we label its left endpoint by $(x,y)$. If $l$ is a vertical link, we label its bottom endpoint by $(x,y)$. Then from the definition of $U$ in fig. \ref{fig:lattice_shear}, 
\be
U S_{l(x,y)}=  S'_{l'(x+y,y)}  U.
\label{eq:S'}
\ee
where operator $S'$ is defined in fig. \ref{fig:sheared_S}. 
Exploiting equation \eqref{eq:S'}, the expectation value of the trace can then be directly evaluated,
\begin{align}
& \tr[\rho_{\mu} U] \nonumber \\
= &  \sum_{\bm{a}}\big(\prod_i F_{a_i}\big) \tr [\rho_0 \prod_l (S_l^{a_l})^{\dagger} U \prod_j S_j^{a_j}] \nonumber \\
= &  \sum_{\bm{a}} \big(\prod_i F_{a_i}\big)\tr [\rho_0 \prod_l  (S_l^{a_l})^{\dagger} \prod_j S_{j'(x_j+y_j,y_j)}'^{a_j} U ] \nonumber \\
= &  \sum_{\bm{a}} \big(\prod_i F_{a_i}\big) \sum_{s,s'} \frac{\lambda_{s,s'}}{9}\langle \psi_{s,s'}| \prod_l  (S_l^{a_l})^{\dagger} \nonumber\\
& \qquad \qquad \qquad \cdot \prod_j S_{j'(x_j+y_j,y_j)}'^{a_j} |\psi_{s,s'}\rangle 
\label{eq:direct_U}
\end{align}
For the matrix element in \eqref{eq:direct_U} to be non-vanishing, the operator in the middle must form closed loops. This can happen if: 
\begin{itemize}
    \item[(i)] The product over $S$ forms product of $O_v$ operators, and the product over $S'$ form the corresponding product of translated $O_v$ operators: $U O_{v(x,y)} U^{\dagger}=O_{v(x+y,y)}$. Summation over error configurations is then dual to the partition function of the three-state Potts model, where the domain walls of the latter model maps to the closed loops in the decohered toric code model. The Potts degrees of freedom $\chi_p$ live on the plaquettes of the square lattice, with the Hamiltonian:
\be
\tilde{H}=- J(\mu) \sum_{\langle pp'\rangle} \chi_p^{\dagger} \chi_{p'} + \text{h.c.}
\ee
When $\chi_p^{\dagger} \chi_{p'}=\omega^a$ for neighboring plaquettes, there is a domain wall of color $a$ on the link separating $p$ and $p'$. This domain wall carries the following relative Boltzmann weight with respect to the trivial domain wall case: 
$e^{-3\beta J}=F_1/F_0$, indicating:
\be
\beta J(\mu) = \frac{1}{3}\ln \frac{F_0(\mu)}{F_a(\mu)}.
\ee
Thus, only when $F_1=F_2$, can we map the calculation into that of a Hermitian Potts model. Restricting the channel to this case, then the problem of summing over closed loop configurations translates into the partition function of the Potts model $Z_{\text{Potts}}$, whose free energy can give rise to a singularity. However, it does not affect the phase factor of $\tr [U \rho_{\mu} ]$ which is all that matters for chiral central charge. 

\item[(ii)] If in addition to potential contractible loops, the product over $S'$ and $S$ each further forms a non-contractible $W_x^{-1}$ loop (which is the product of the left panel in fig. \ref{fig:Kraus} over the $x$-direction, or the product of the left panel in fig. \ref{fig:sheared_S} over the $x$-direction). For each such horizontal non-contractible loop, the contribution to $\tr[\rho_{\mu} U]$ is 
\be
\frac{1}{3} F_0^{M}\left[ (F_1/F_0)^L+(F_2/F_0)^L\right].
\ee
The two separate terms come from the fact that both $W_x$ and $W_x^{-1}$ are possible, and we have taken into account  the summation over $s, s'$ and used $\sum_{s,s'} \lambda_{s,s'}=3$ from \eqref{eq:iteration}. 

\item[(iii)] The product over $S$ operators forms the non-contractible $W_y$, while the product over $S$ forms $U W_y U^{-1} \sim \omega W_x W_y$ up to contractible loops which evaluate to $1$ in $\rho_0$. Each matrix element thus gives
\begin{align}
& \sum_{s,s'}\frac{\lambda_{s,s'}}{9}\langle \psi_{s,s'}|W_y^{\dagger} \omega W_x W_y|\psi_{s,s'}\rangle \nonumber\\
= & \sum_{s,s'}\frac{\lambda_{s,s'}}{9} \omega^{s-1}=\frac{1}{3}\omega^{-2} = \frac{\omega}{3}.
\end{align}
Consequently, such a vertical non-contractible loop contributes to the expectation value of $U$ as
\be
\frac{\omega }{3} F_0^{M}\left[ (F_1/F_0)^L+(F_2/F_0)^L)\right].
\label{eq:omega_F}
\ee
\end{itemize}
Only in case (\rm{iii}) when the error configuration includes vertical non-contractible loop $W_y$, can a complex phase factor be obtained. However, the ratio between the Boltzmann factors with a $W_y$ and without a $W_y$ is proportional to the factor $(F_1/F_0)^L = (F_2/F_0)^L$. By construction of the decoherence channel, $(F_1/F_0)\in [0,1]$. Therefore, in the thermodynamic limit, the influence of non-contractible loops goes to zero unless $\mu=\infty$ and $F_1=F_2=F_0.$ In the latter case, case (\rm{iii}) in \eqref{eq:omega_F} gives $2\omega F_0^M/3$ for the two nontrivial $W_y$ loops, and combining with the trivial loop case, we have $(1+2\omega)F_0^M/3$. The factor of $F_0^M$ corresponds to the Potts partition function, and $(1+2\omega)/3=\ii/\sqrt{3}$ whose phase angle is $\pi/2$, leading to $c^-_{\infty}=2$, consistent with the calculation in the previous section \eqref{eq:c_infty}. 

To summarize, $\tr[U\rho]$ does not contain information to detect chirality beyond fixed points due to the fact that the chirality information lives in the non-contractible error loops. The latter always have a subdominant contribution compared with the configuration without such loops. Even if one attempts to generalize this expression to its higher Renyí version, we still expect it to be blind to the correct chirality transition point.

\clearpage

\bibliography{ref.bib}

\end{document}